# Legends of *Nature* and Editors-in-chief

Zhiwen Hu[1]*, Chuhan Wu[2]*, Zhongliang Yang[2]*, Yongfeng Huang[2†]

[1]School of Computer and Information Engineering, Zhejiang Gongshang University, Hangzhou 310018, China.
[2]Beijing National Research Center for Information Science and Technology, Tsinghua University, Beijing 100084, China.
[†]Correspondence and requests for materials should be addressed to yfhuang@tsinghua.edu.cn.

**Author contributions**

Z.W.H., C.H.W. and Z.L.Y. contributed equally to this work. Z.W.H. and Y.F.H. were involved in the conceptual design of the study. Z.W.H., C.H.W. and Z.L.Y. performed the data analyses. Z.W.H., C.H.W., Z.L.Y. and Y.F.H. wrote the manuscript.

**Competing interests**

The authors declare no competing interests.

**Acknowledgments**

The authors thank Mr. Yubo Chen and Jiahao Qin for corpus construction. This work was supported in part by the National Key Research and Development Program of China under Grant 2018YFC1604002 and the National Natural Science Foundation of China under Grant U1705261, Grant U1836204, Grant U1536201 and Grant U1636113.

**Data and materials availability**

The dataset analyzed during the current study is available in the Github repository, https://github.com/wuch15/Time-for-Editors-in-Chief-of-Nature.

## Abstract

This year marks the 150th celebration of *Nature*. However, the understanding of the way the army of unsung editors-in-chief has strengthened and enriched the integrity and quality of the journal under the umbrella of its original mission remains nominal rather than substantial. This paper scrutinizes the chief vehicle guided by *Nature*'s doctrine with regard to the ways it has conflicted with the advancement of both science and social progress. We first recast quantitative spatiotemporal analysis on the diachronic discourse of *Nature* since its debut, which promises to articulate the unfolding chronological picture of *Nature* on a historical time scale, and pinpoint overdue corrective to the strongly-held but flawed notions on editors-in-chief of *Nature*. Our findings strongly indicate that the army of editors-in-chief have never met with their fair share of identification, and they took on the challenge guided by *Nature*'s doctrine with extraordinary polymath, unparalleled enthusiasm and diverse characters.

## Introduction

The simple title *Nature*, embracing all in a single word, was appropriately first chosen by Sir Joseph Norman Lockyer when he founded this "weekly illustrated journal of science" in 1869, which was initially launched from a seminal idea in 1868 (Deslandres, 1919; Roy Malcolm MacLeod, 1969d; "'Nature' and the Macmillan Company," 1943). This year marks the sesquicentennial celebration of *Nature*. It is high time to reconsider the unsung heroes in the editorial chair of this leading scientific journal, who were imbued with faith in "the salvation of man through science"(Werskey, 1969). However, the understanding of the ways the army of unsung editors-in-chief has strengthened and enriched the integrity and quality of the journal under the umbrella of its original mission remains nominal rather than substantial. Similar to the centennial festschrift (Werskey, 1969), this paper scrutinizes the chief vehicle guided by *Nature*'s doctrine with regard to the ways it has conflicted with the advancement of both science and social progress.

## Meta-analysis and evidence synthesis

The Google Books Corpus (GBC)(Michel et al., 2011) is a unique linguistic landscape that benefits from centuries of development of rich grammatical and lexical resources as well as cultural contexts. SentiWordNet (Esuli, Sebastiani, & Moruzzi, 2006) is a large-scale sentiment lexicon in which each WordNet synset is labelled with three real-valued scores to describe how objective, positive and negative the terms are in the synset. The John Maddox Corpus (JMC) is a collection of Sir John Royden Maddox's publications in *Nature* and *Science*, which have been converted into textual formats to facilitate subsequent analysis via natural language processing (NLP).

To characterize the scientific fame of specific persons, the GBC is employed to compute $N_i$, *i.e.*, the appearance frequency of a personal name in the year *i*, which represents how often the name is mentioned over time (Michel et al., 2011). In addition, the $N_i$ scores that are non-zero before birth are removed to filter potential noisy records. The GBC covers the data logs from 1800 to mid-2009. To orchestrate the fame scores before and after 2008, the Google Scholar search engine is utilized by querying personal names from the logs in each year. Denote the retrieval results in year *i* as $G_i$; then, the appearance frequency $\alpha_i$ in Google Scholar is formulated as

$$\alpha_i = \frac{G_i}{M_i},$$

where $M_i$ is the number of logs in Google Scholar in the year *i*. Scientific fame $N_i$ $(i > 2008)$ is computed as

$$N_i = \frac{\alpha_i}{\alpha_{2008}} N_{2008} .$$

To illustrate the prosperity of *Nature* and several other learned journals, the number of publications indexed by Web of Science over the years is used. To filter possible incorrect or mismatched logs, the retrieval results before the year of each journal's establishment are removed.

To extract representative sentiment words from the JMC, SentiWordNet is employed to compute the positive score $s_p \in [0,1]$ and the negative score $s_n \in [0,1]$ of each word in texts. The overall sentiment score $s$ is computed as

$$s = s_p - s_n .$$

Because negation words usually influence the sentiment polarities, the signs of the sentiment scores of words modified by negation words are converted. Additionally, because words with weak sentiment orientations may be more difficult to correctly identify, words with $|s| \leq 0.5$ are filtered and others are recognized as sentiment words. All positive and negative words are ordered by their frequency, and they are portrayed as word clouds in Figure 3.

To extract key topic words of John Maddox, the Latent Dirichlet Allocation (LDA)(Blei, Ng, & Jordan, 2003) algorithm is employed. To show their semantic similarities, pre-trained word embeddings are used to represent these words in a vector space. For visualization, the t-SNE algorithm is used to project the high-dimensional word vectors into a plane, and the results are shown in Figure 4.

To analyze the semantic evolution of the word "computer" over the years, the 5-grams of GBC in each year is used to train word embedding vectors from their contexts via word2vec (Mikolov, Chen, Corrado, & Dean, 2013). After obtaining the distributed representation $\vec{\boldsymbol{w}}_i$ of each word $w_i$, the semantic similarity $r_{ij}$ of a pair of words $(w_i, w_j)$ is computed as

$$r_{ij} =< \vec{\boldsymbol{w}}_i, \vec{\boldsymbol{w}}_j >= \frac{\vec{\boldsymbol{w}}_i \cdot \vec{\boldsymbol{w}}_j}{||\vec{\boldsymbol{w}}_i|| \cdot ||\vec{\boldsymbol{w}}_j||} .$$

The semantic similarity of the word "computer" with each other word in GBC is computed for each year, and the top similar words from 1858 to 2008 are illustrated in Figure 2 (10 words/10 years).

## Timeline of Editors-in-chief

*Nature*, Charles Darwin's favorite journal(Witkowski, 2016), is devoted to all the sciences, alongside *The Lancet* (founded in 1823) and *Science* (founded in 1880). On 4 November 1869, *Nature* launched its first epoch-making issue ("The First Issue," 1969) with an ambitious marching order from the English Romantic poet William Wordsworth's poetry (R. M. MacLeod, 1969):

"*To the solid ground of Nature trusts the mind which builds for aye*."

The initial doctrine of *Nature* was "to promote science that was accessible, but that did not involve the public in scientific discussions" (Baber, 2008; "Selections from the Letters of Sir Norman Lockyer," 1969).

The chronological account of the dedicated editors-in-chief of *Nature* is a faithful mirror of this renowned journal (Figure 1). Sir Joseph Norman Lockyer (17 May 1836 – 16 August 1920) was the founding editor of *Nature* from 11 November 1869 to 6 November 1919 (Arch. Geikie, W. T. Thiselton-Dyer, T. E. Thorpe, William A. Tilden, Clifford Allbutt, T. G. Bonney, 1920; Baber, 2008; Lockyer, 1922; “Sir Norman Lockyer, 1836–1920,” 1936). He is credited as the father of gas helium, along with French scientist Pierre Janssen. Lockyer is regarded as “one of the great men of science of England and one of the greatest astronomers of all time” (Deslandres, 1919; Fowler, 1920; “Sir Norman Lockyer, 1836–1920,” 1936). Despite financial difficulties that continued for several years (Roy Malcolm MacLeod, 1969b), Lockyer took the helm, and his fame and reputation spread. In 1889, Sir Richard Arman Gregory (29 January 1864 – 15 September 1952) returned to the Royal College of Science as computer for the Solar Physics Committee and assistant to Norman Lockyer (Hartley, 1953)(Figure 2). Later, at *Nature*’s Jubilee, Lockyer officially relinquished his editorship to Gregory (Roy Malcolm MacLeod, 1969a), who had been de facto editor since 1912 until his retirement in December 1938 at the age of 75 (Gregory, 1939a, 1939b; W. N. McClean, 1953; Werskey, 1969). The day before WWII broke out, he had been elected president of the British Association for the Advancement of Science (BAAS).

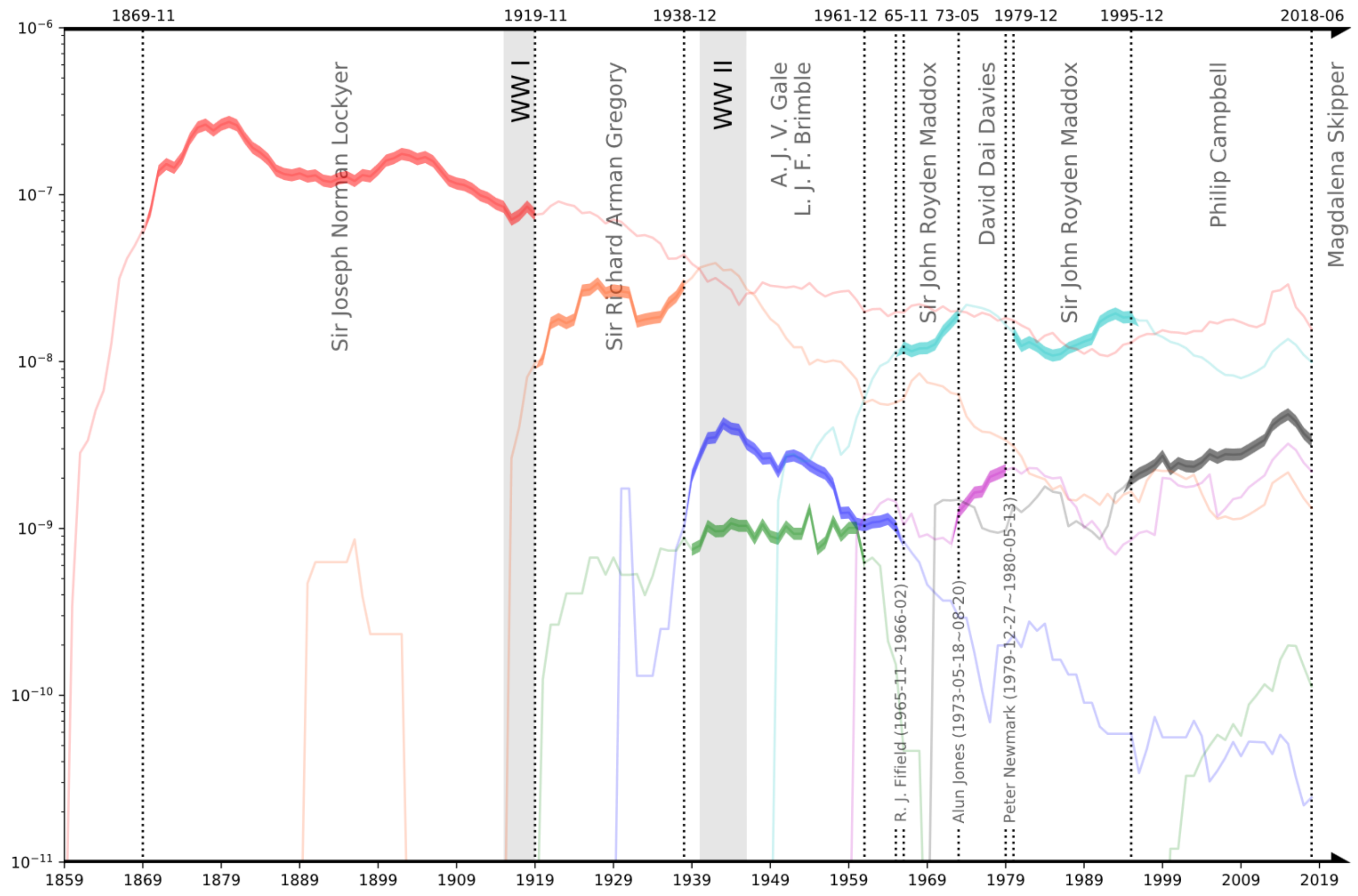

**Fig. 1.** The chronological account of the dedicated editors-in-chief of *Nature* is a faithful mirror of this renowned “weekly illustrated journal of science”. The scientific fame curves between 1859 and 2008 are facsimiled according to the *n*-gram frequency in the Google Books Corpus, and those after 2008 are obtained from the Google Scholar search engine due to the coverage of the Google Books *n*-gram database. Our findings indicate that Sir Joseph Norman Lockyer (red), Sir Richard Arman Gregory (orange), Lionel John Farnham Brimble (blue), Arthur J. V. Gale (green), Sir John Royden Maddox (cyan), David Dai Davies (magenta) and Philip Campbell (black) enjoyed fame and reputation during their tenures, and the collective memory of these unsung journalists has both short-term and long-term components.

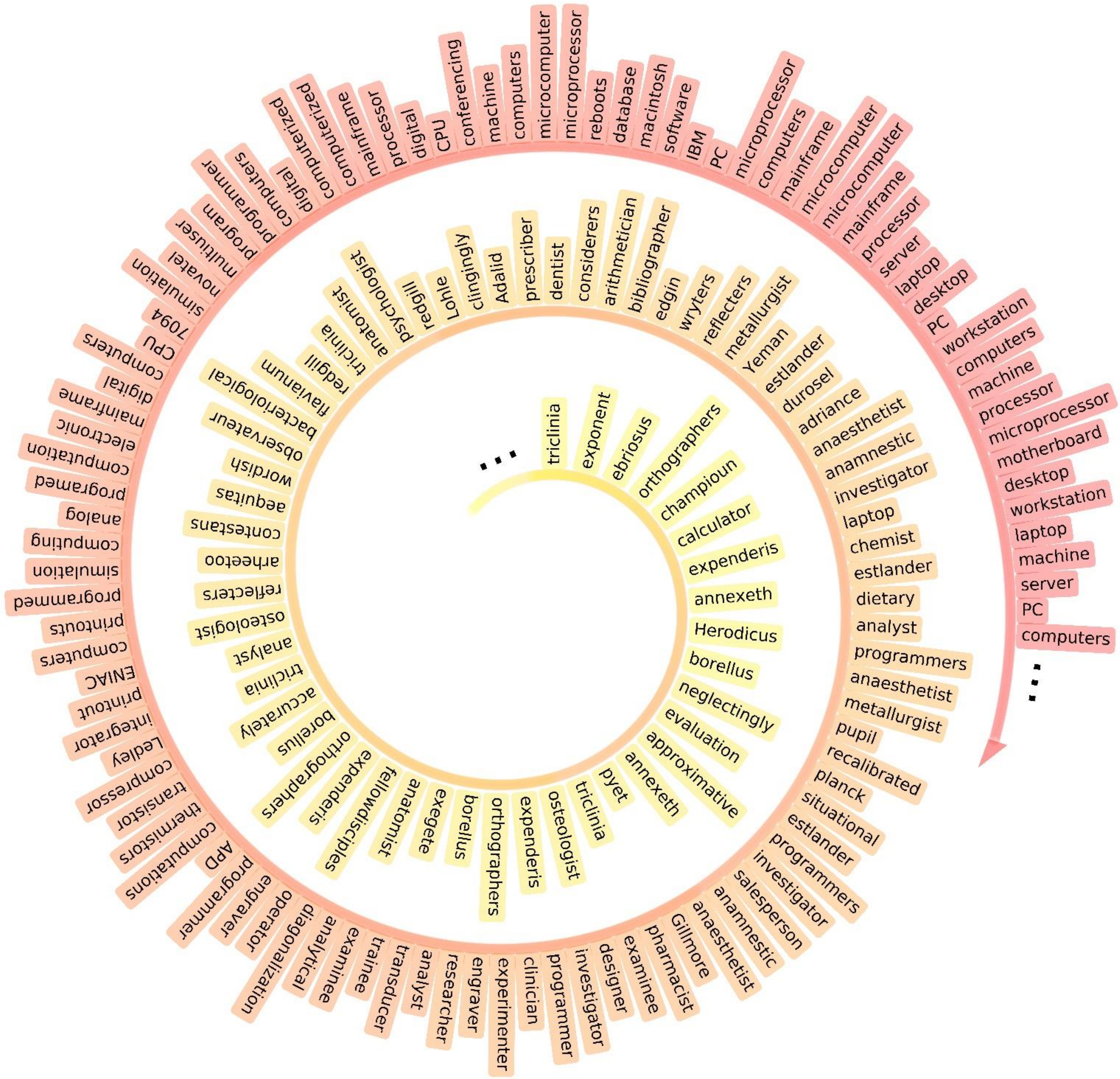

**Fig. 2.** Spiral projection of words similar to 'computer'. According to the historiographical narratives, Sir Richard Arman Gregory and John Royden Maddox were computers in their early careers. Similarities of words are computed by the cosine distances of word embedding vectors pre-trained on the Google Books Corpus in different years. In the early years, 'computer' was used as an occupation like many similar words. In later years, its meaning focused on the computer that we are familiar with today.

With the outbreak of war, Richard Gregory handed the editorship over to Lionel John Farnham Brimble (16 January 1904 - 15 November 1965) and Arthur J. V. Gale (1895 – 4 September 1978). As co-editors, Brimble and Gale carried *Nature* through WWII into an era of rapidly expanding science. At the end of 1961, Gale retired from the joint editorship, leaving Brimble as the sole editor. On 15 November 1965, Brimble died at his desk in London late at night. Assistant Editor R. J. Fifield temporarily took on Brimble's duties until the appointment of Sir John Royden Maddox (27 November 1925 – 12 April 2009) in February 1966. Maddox served as editor until 18 May 1973. Dr. Alun Jones, the Deputy Editor, temporarily took on Maddox's duties until the end of August. Soon after, David Dai Davies (born in 1939) took over the editorship on 20 August 1973. Davies resigned on 27 December 1979, and Dr. Peter Newmark, the Deputy Editor, took over the interim editorship as Acting Editor of *Nature* until 13 May 1980. At that point, Maddox resumed the position of editor and retained this position until 8 December 1995. After Maddox resigned, he became an Editor Emeritus of *Nature*. On 8 December 1995, Philip Campbell succeeded Maddox in the editorial chair. After Campbell stepped down from his editorship on 1 July 2018, Magdalena Skipper became the first woman and life scientist to hold this post at *Nature* (Table 1). Before that, she was a senior editor of *Nature* and chief editor of *Nature Reviews Genetics* and *Nature Communications*.

**Table 1** Chronological List of Editors-in-chief of *Nature*.

| # | Name | Tenure | Known for |
|---|---|---|---|
| 1 | Sir Joseph Norman Lockyer | 11 Nov. 1869 – 6 Nov. 1919 | astrophysicist, spectroscopist, meteorologist, archaeological astronomer |
| 2 | Sir Richard Arman Gregory | 6 Nov. 1919 – Dec. 1938 | astrophysicist, educationist, Hellenist |
| 3 | Arthur J. V. Gale | Jan. 1939 – Dec.1961 | agronomist |
| | Lionel John Farnham Brimble | Jan. 1939 – Nov. 1965 | botanist, theatre criticist |
| 4 | R. J. Fifield* | Nov. 1965 – Feb. 1966 | assistant editor |
| 5 | Sir John Royden Maddox | Feb. 1966 – 18 May 1973 | theoretical physicist, cosmologist, biologist |
| 6 | Alun Jones* | 18 May 1973 – 20 Aug. 1973 | deputy editor |
| 7 | David Dai Davies | 20 Aug. 1973 – 27 Dec. 1979 | geophysicist, seismologist |
| 8 | Peter Newmark* | 27 Dec. 1979 – 13 May 1980 | deputy editor |
| 9 | Sir John Royden Maddox | 13 May 1980 – 8 Dec. 1995 | theoretical physicist, cosmologist, biologist |
| 10 | Philip Campbell | 8 Dec. 1995 – 28 Jun. 2018 | physicist |
| 11 | Magdalena Skipper | 1 Jul. 2018 – Present | geneticist, molecular biologist |

*Note: R. J. Fifield, Alun Jones and Peter Newmark held the post of interim editor-in-chief for different reasons.

## Same mission, different character

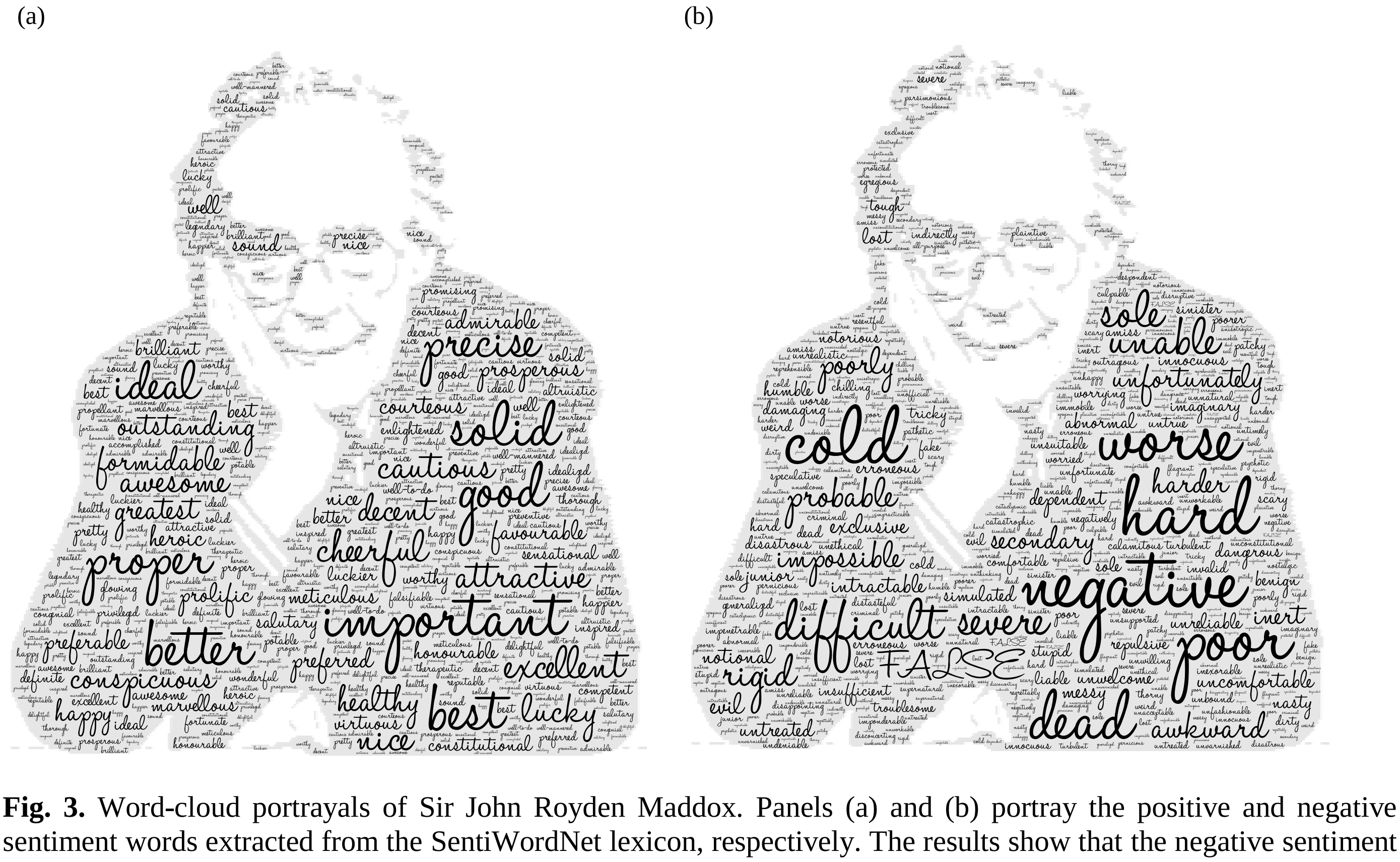

**Fig. 3.** Word-cloud portrayals of Sir John Royden Maddox. Panels (a) and (b) portray the positive and negative sentiment words extracted from the SentiWordNet lexicon, respectively. The results show that the negative sentiment of these editorials is determinant. As a case in point, Maddox penned serial hard-hitting editorials to scold molecular biologists regarding the philosophy of molecular biology. (For an interactive version of these graphics, visit: https://ngnlab.cn/2019/time-for-editors-in-chief)

All of these dedicated editors, different in character, repeatedly put *Nature* on the right track in the past 150 years, and the past pages, whatever merits and demerits, deserve credit for recognizing this. As the longest-serving editor-in-chief, Lockyer was preeminently successful in securing "the identification of the personalities" of scientific workers and advanced the scientific opinion of the day with his journal, as did his successful successor ("Sir Norman Lockyer and the Editorship of '*Nature*,'" 1929). Lockyer had enough charm to encourage an army of eminent advocators to work for *Nature*, but his ruthlessness cost him in friendship. For example, in May 1870, John Brett complained that Lockyer has not given him the appropriate priority, although he had provided free astronomy coverage to Lockyer (Roy Malcolm MacLeod, 1969c). Gregory was a perennially cheerful, generous and friendly person. Brimble was notoriously rough, exemplified by his frequent rebukes of John Maddox's criticisms of *Nature* (Maddox, 1995). Dr. David Davies, a geophysicist affiliated with the Massachusetts Institute of Technology in the United States, has enjoyed his leadership with a firm, affable but principled character (Bondi, 1980).

This year also marks the 10$^{th}$ death anniversary of Sir John Royden Maddox (Byam Shaw, 2009; Campbell, 2009; "Editor of Nature," 1980a; "Editor of Nature," 1980b; "New editor is appointed at Nature," 1995; Gratzer, 2010; Maddox, 1995; Wade, 2009; Watts, 2009), formerly editor-in-chief of *Nature* and occasional editor of the 100- and 125-year anniversary issues of *Nature*. As a peerless science correspondent with an international outlook, he first established editorial offices in the United States, France, Germany and Japan. As a computer at the University of Manchester (Figure 2), Maddox briefly teamed up with Alan Turing (Gratzer, 2009), although he was not mentioned (Maddox, 1985). Maddox had a unique personal substance: he was red-blooded, adventurous, and prodigious in memory (Figure 3).

As a case in point, Maddox's most ambitious venture was to split *Nature* into the portfolio of *Nature*, *Nature Physical Science* and *Nature New Biology* ("Nature Subscriptions," 1973)(Figure 4). That short-lived experiment may have resulted in his departure in 1973 (Gratzer, 2010). When *Nature* returned on track in 1974, Benjamin Lewin, left *Nature New Biology*, where he was the moderator, and founded the leading journal *Cell* (Witkowski, 2016). In 1980, Maddox returned to his editorial chair. After his retirement at the age of 70, he became an honorable Editor Emeritus of *Nature*.

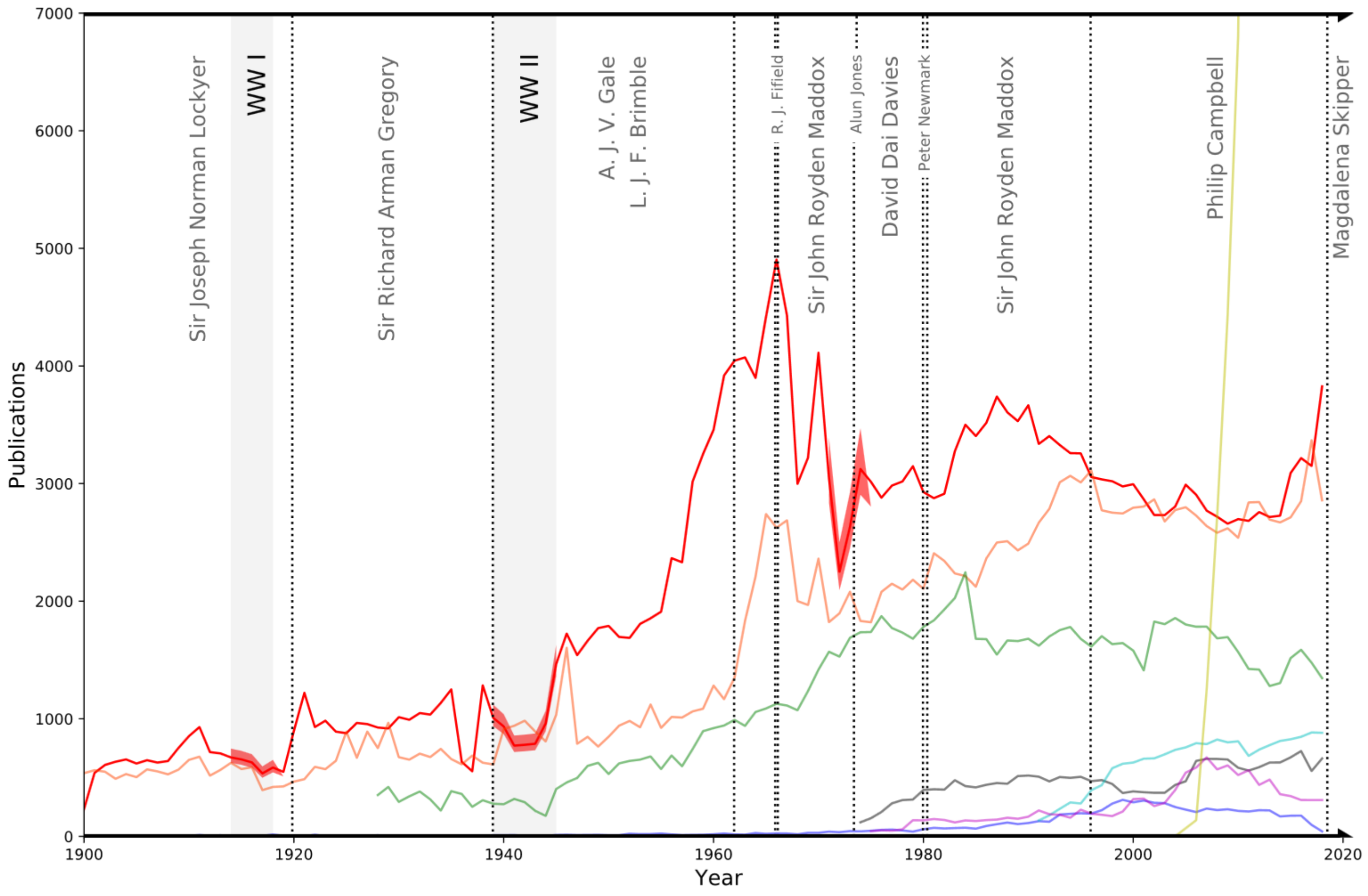

**Fig. 4.** Metadata of Web of Science facsimiles of the patterns of issued publications in *Nature* and control groups of learned journals. In *Nature*'s first fifty years, the founding editor Lockyer and his assistants produced 2,600 weekly publications in 103 volumes. As a cutting-edge weekly journal, *Nature* (red) has survived through the World Wars, along with *Science* (orange) and *The Lancet* (blue). International conflicts significantly inhibited the issued publications of *Nature*, *Science* and *The New England Journal of Medicine* (green). Notably, substantial pattern shifts are observed before and after each editor-in-chief turnover of *Nature*. This finding indicates that maintaining a relatively stable

editorial team is beneficial to the development of a scientific journal. For example, *Nature* experienced dramatic fluctuations between 1971 and 1974, which coincided with John Maddox's venture of splitting *Nature* into the portfolio of *Nature*, *Nature Physical Science* and *Nature New Biology*. When *Nature* got back on track, the molecular biologist Benjamin Lewin left *Nature New Biology* and founded *Cell* (gray) in January 1974. In 1975, David F. Horrobin, as a founding editor, established a non-peer-reviewed medical journal, *Medical Hypotheses* (magenta), to challenge the peer-review policy. In April 1992, Geoffrey North, deputy biology editor of *Nature*, left *Nature*, where he had worked for more than 11 years. He was then appointed the editor-in-chief of the biweekly peer-reviewed journal *Current Biology* (cyan), which was established in 1991. As the first multidisciplinary open-access journal, *PLoS One* (yellow) debuted in 2006 and culminated as the world's largest journal in 2013.

## Diverse topics, different voices

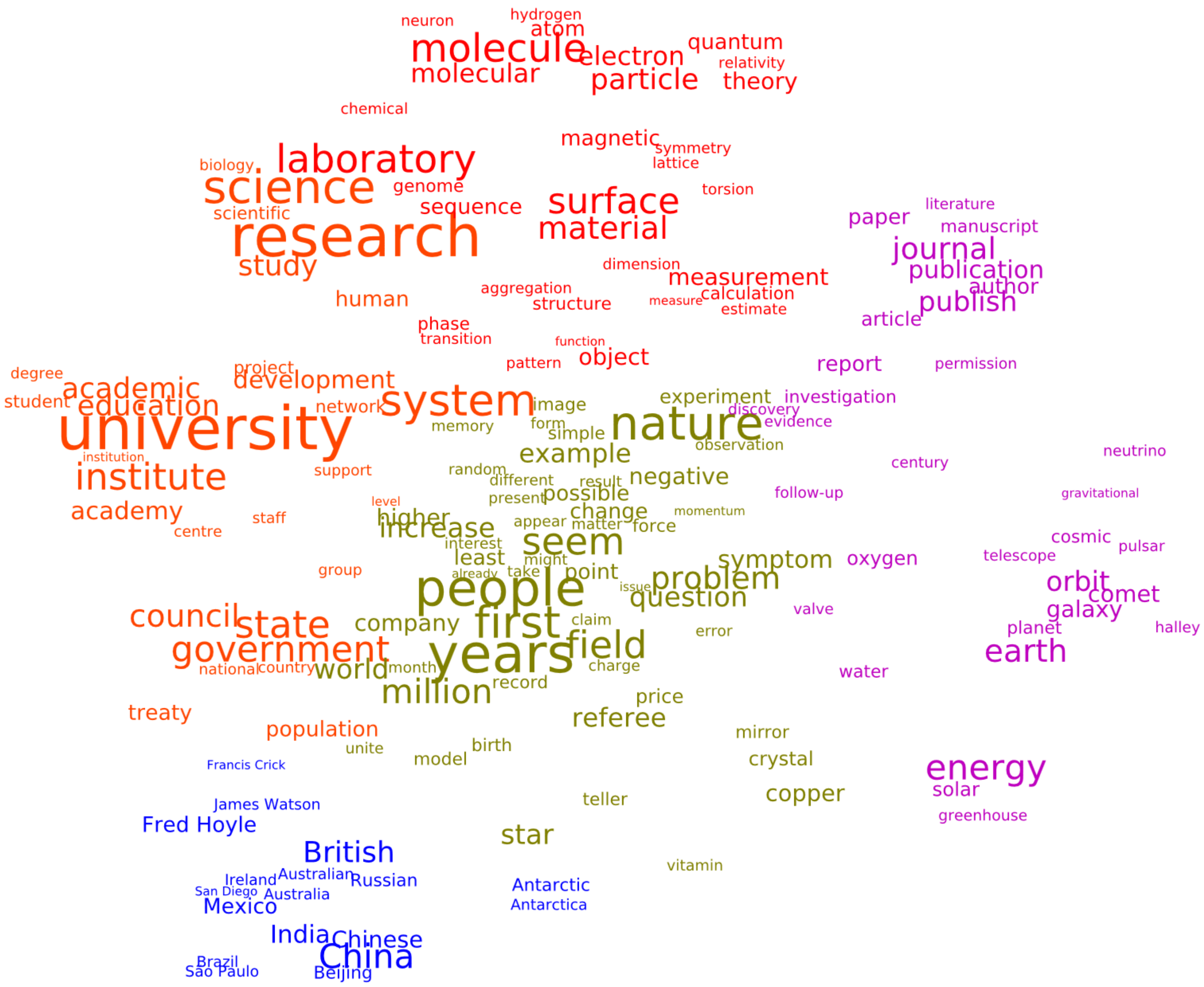

**Fig. 5.** Topic-word projection of Sir John Royden Maddox. The topic words are extracted by the latent Dirichlet allocation algorithm and are visualized by projecting the word-embedding vectors via t-SNE. *Nature* has had many eminent advocates, including Fred Hoyle, James Watson, Francis Crick, and Charles Darwin.

With the vitality of science in all walks of life, no matter how experienced the editor-in-chief is, enormous challenges must be addressed. As one of the most prolific editors of *Nature* with unequaled enthusiasm throughout two long tenures, Maddox penned approximately 600 editorials on prescient developments in the liveliest areas, ranging from the Big Bang origin of the universe to the Non-Proliferation Treaty, from molecular biology to AIDS, from the British biologist Rupert Sheldrake's books to the Hay-on-Wye festival, from scientific misconduct to the passive voice in scientific writing, and from the abortion policy of the United States to the eugenics law of China (Figure 5). In his valedictory article as the editor of *Nature* (Maddox, 1995), John Maddox warned,

"*Misconduct of some overt kind is on the rise. We all know why that is. Reputations rest on publications as never before, as do promotions and research grants*."

Maddox even extended his commentaries to *Science*, including the most controversial debate – the Benveniste affair (Gratzer, 2009; Maddox, 1988).

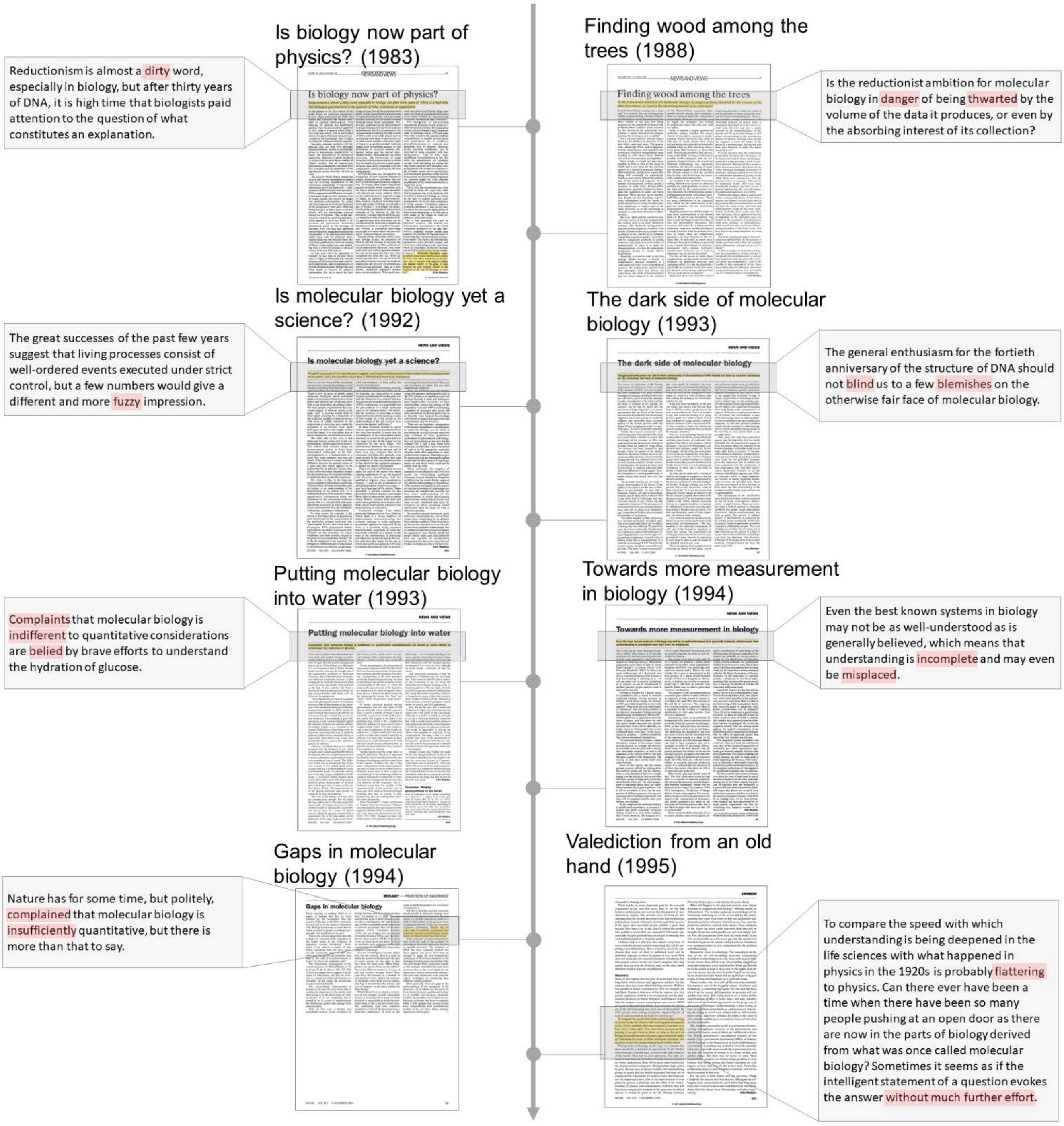

**Fig. 6.** Maddox's eight-combo critiques upon molecular biology. Sir John Royden Maddox penned a series of hard-hitting editorials from 1983 to 1995 to scold molecular biologists regarding the philosophy of molecular biology. These eight-combo critiques, which are exemplified here by highlighted content with negative sentiments, invoked intensive arguments among scientists.

Unfortunately, after 22 eventful years, Maddox was labelled a "man with a whim of iron" (Gratzer, 2009; Maddox, 1995) and "a correspondent in molecular biology" (Witkowski, 2016). Maddox penned a series of hard-hitting editorials from 1983 to 1995 to scold molecular biologists regarding the philosophy of molecular biology (Figure 6). These eight critiques invoked intensive arguments among scientists (Braaten, 1993; Kell, 1999; Lortie, 1993; Segel & Tyson, 1992; Shlesinger, Zaslavsky, & Klafter, 1993; Spiller, Wood, Rand, & White, 2010; Steinmetz, 1993; Wheatley, 1993). Thomas L. Clarke argued against Maddox's reductionism (Clarke, 1984):

"*If biology is in principle reducible to physics, then so is psychology*."

Lee Segel and John J. Tyson defended Maddox's denouncements with rapid correspondence (Segel & Tyson, 1992), as did M. F. Shlesinger and his colleagues (Shlesinger et al., 1993). Douglas Braaten argued against John Maddox's reductionism, broadly understood as a belief in the explanation of biological phenomena wholly in physical and chemical terms (Braaten, 1993). Robert Lortle valued the nonlinear quantification effort in open dynamic systems in the molecular biology field, but he also argued against the reductionist approach in molecular biology (Steinmetz, 1993). Denys N. Wheatley dismissed Maddox's denunciations and Segel and Tyson's echoes (Wheatley, 1993). In 1999, Douglas B. Kell agreed with Maddox's denouncements but refuted the grounds of his argument (Kell, 1999).

On eugenics, in 1904, Joseph Norman Lockyer published an editorial that abridged from the note read before the Sociological Society by Sir Francis Galton ("Eugenics: Its Definition,Scope and Aims," 1904; Galton, 1904), who is known as the father of eugenics and half-cousin of Charles Darwin (Comfort, 2019). Twenty year later, Richard Gregory, a former vice-president of Marie Stopes' Society for Constructive Birth Control and Racial Progress (Werskey, 1969), published *Nature*'s first pro-eugenics editorial in May of 1924 (MacBride, 1924). But seventy-one years later, Maddox proposed a critical review of the Chinese Maternal and Infant Health Care Law together with David Swinbanks (Maddox & Swinbanks, 1995), who is known as the founder of the Nature Index.

## Conclusions and implication

In retrospect, the army of editors guided *Nature* from a gossip sheet ("What is Nature for?," 1973) to a learned journal after eventful years, and their epoch-making efforts should be fairly recognized. In *Nature*'s first fifty years, Lockyer diligently penned 66 editorials in *Nature*. Unfortunately, Lockyer was guilty of abusing his editorship, exemplified by signing manuscripts for intimates, even himself, and he relinquished the editorship (Roy Malcolm MacLeod, 1969c). In the same token, Brimble and Gale were also accused and relinquished editorship to Rainald Brightman for 27 years (Witkowski, 2016). Further, they were reproved for the autocratic review process ("Effects of Sexual Activity on Beard Growth in Man," 1970; Witkowski, 2016). Maddox continued to send some manuscripts out for peer review but rejected and accepted others on his own authority. This arbitrary operation of an anonymous article, entitled "*Effects of sexual activity on beard growth in man*," aroused great consternation ("Effects of Sexual Activity on Beard Growth in Man," 1970). In the article, he even claimed:

> "*The identity of the author of this communication has been suppressed for reasons which may be self-evident, but the author, whose work has been vouched for by a colleague, has answered a number of questions raised by a referee.*" (p. 870)

However, according to Geoffrey North (North, 2004), formerly deputy biology editor of *Nature* from 1981 to 1992, *Nature* finalized the command of computers in April 1992. Before that, piled-up manuscripts were sent by ordinary mail to potential reviewers without knowing whether they were able to assess the manuscripts. Defensibly, publish-or-perish is always the real dilemma for editors. Beyond that, rational critiques were dedicated to shaping *Nature*'s future and that of its rivals and other scientific journals.

At critical moments, scientific journals must make informed choices (Aronson et al., 2019). In 1975, David F. Horrobin, as a founding editor, established a non-peer-reviewed medical journal *Medical Hypotheses*, to challenge the peer-review policy. Thirty years after the first congress on peer-review policy, the majority of journals made different choices. With the growth of computers and the Internet, scientific journals have experienced a substantial transition from ink-on-paper journals to electronic ones. In fact, Maddox was cautiously optimistic about e-journals (Maddox, 1992):

"*Electronic journals may already have arrived, but their management remains a problem for the future*."

As forewarned, *The Online Journal of Current Clinical Trials*, the first peer-reviewed scientific journal published online without a print version, survived only four years. More recently, dozens of multidisciplinary open-access journals have engaged unique niches in scientific publications, in line with the founding missions of barning timely sound science regardless of perceived novelty. But high publication fee, large volume and impact factor decline are gradually undermining their early momentums, including the trailblazers *PLoS One* and *Scientific Reports*.

*Nature* should take an open mind to appreciate modest introspections and rededications to celebrate its sesquicentenary. Just as the centennial festschrift remarked (R. M. MacLeod, 1969):

"*Nature will in the years ahead seek new and ingenious ways of recapturing more of the old directness*".

The farther back we can look, the farther forward we can see. This is a positivist tenet, not merely for learned journals such as *Nature* but for the vitality of science and the promotion of social progress.